# Probability that a chromosome is lost without trace under the neutral Wright-Fisher model with recombination

Badri Padhukasahasram[1]

[1]Section on Ecology and Evolution, University of California, Davis

Address for correspondence: 4341 Genome Center, 451 Health Sciences Drive, University of California, Davis, California, U.S.A. 95616

Email: pkbadri@yahoo.com

**Abstract**

I describe an analytical approximation for calculating the short-term probability of loss of a chromosome under the neutral Wright-Fisher model with recombination. I also present an upper and lower bound for this probability. Exact analytical calculation of this quantity is difficult and computationally expensive because the number of different ways in which a chromosome can be lost, grows very large in the presence of recombination. Simulations indicate that the probabilities obtained using my approximate formula are always comparable to the true expectations provided that the number of generations remains small. These results are useful in the context of an algorithm that we recently developed for simulating Wright-Fisher populations forward in time.

**Introduction**

The Wright-Fisher model refers to a popular stochastic model of reproduction in the area of population genetics (see Fisher 1930, Wright 1931). The variant that is used here allows for recombination and considers a population of constant size with $N$ organisms (with a total of $2N$ chromosomes) in each generation. Each organism consists of 2 chromosomes and lives for exactly 1 generation. Everyone dies right after the offspring are made and thus generations are non-overlapping. The reproduction scheme is such that one can think of each individual in the next generation as receiving two chromosomes, each one selected randomly and with replacement from the chromosomes present among the parents. The next generation, thus, will be $2N$ chromosomes, sampled with replacement from the $2N$ in the previous generation. Although the above description is highly idealized, the model manages to capture the essence of reproduction in real populations and its assumptions can be relaxed to make it more realistic.

Chromosomes that occur within the same organism will be referred to as homologous chromosomes or homologs of one another. All chromosomes consist of $L$ basepairs of DNA. Recombination breaks can occur between consecutive basepairs and a chromosome can have at most 1 recombination per generation every time it gets sampled. After sampling each chromosome, we determine if it recombines based on the probability of recombination. If $r$ denotes the per-generation per-sequence recombination rate, this probability equals $1 - e^{-r}$. If there is no recombination, then the sampled chromosome is simply copied to the next generation. If a recombination event happens, we create a mosaic chromosome in the next generation by choosing an integer $i$ uniformly between $1$ and $L - 1$, and by copying from $1^{st}$ basepair to $i^{th}$ basepair from the sampled chromosome and from $i + 1^{th}$ basepair to $L^{th}$ basepair from its homolog. The new chromosome, thus created, will be referred to as a descendant of the sampled chromosome. Any chromosome created in the subsequent generations by sampling the descendants will also be referred to as descendants of the initial chromosome. Any chromosome in the subsequent generations that contains material from the initially sampled chromosome will be called a copy of that chromosome.

The long-term behaviour of the Wright-Fisher model without recombination has been well studied. For example, if we consider a single chromosome at generation 0, then with probability *1*, we know that the number of copies of this chromosome will eventually reach *0* or *2N*. The probability that a chromosome will eventually be lost is equal to: *1 - 1/2N* and the expected time to loss is approximately *2ln(2N)* generations (Kimura and Ohta 2001). If we start with *i* chromosome copies initially, the expected number of copies in all subsequent generations is also *i*. This property is called *constancy of expectation*. On the other hand, if recombination is included in the model, it becomes considerably less tractable because the state space of the resulting Markov Chain is enormous. In the retrospective coalescent framework (Kingman 1982), there exist efficient computational techniques to trace the ancestry of a sample when recombination is considered and important properties of the ancestral recombination graph were established in Griffiths 1981 and Hudson 1983. Discussions of the coalescent with recombination can be found in the recent monographs in Durret (2002), Hein et al. (2005) and Wakeley (2008). Here, I am interested in the short-term behaviour of the Wright-Fisher model in the presence of recombination and this has not been studied before. With the recent growth in the power of computers, it is now possible to efficiently simulate this process with recombination, forward in time, even for large datasets (e.g. see Hoggart et al 2007, Kimmel and Peng 2005, Padhukasahasram et al. 2008, Hernandez 2008). The results that follow are useful when conducting such forward simulations.

In Padhukasahasram et al. 2008, we introduced a novel algorithm for forward simulation of populations evolving under the neutral Wright-Fisher model with recombination. In that article, we describe 2 different classes of chromosomes in the current population, which do not contribute any material to the future population that exists after $k$ generations from now. These are:

i) A chromosome that loses all its descendants within the next $k$ generations without any of the homologous chromosomes of either the chromosome or its descendants having undergone recombination.

ii) A chromosome whose homologous chromosome recombines in the first generation but both of them lose all their descendants without any of the homologous chromosomes of their descendants having undergone any recombination in the last $k – 1$ generations.

Assuming the per-generation per-sequence recombination rate to be $r$, let $L(k, r, 2N)$ denote the probability that a chromosome belongs to either one of these 2 categories. It is easy to see that $L(k, r, 2N)$ gives a lower bound for the probability that a chromosome is non-ancestral to the future population. This is because there also exist other categories of chromosomes that don't leave behind a trace after $k$ generations. As such, analytically calculating the exact probability (denoted by $NA(k, r, 2N)$) that any chromosome will be non-ancestral to the future population, appears to be a tedious problem when the model includes recombination because the number of possibilities to consider, grows very large. However, it is easy to establish the bounds: $L(k, r, 2N) <= NA(k, r, 2N) <= L(k, 0, 2N)$.

In this article, I will first present a formula for calculating the value of $L(k, r, 2N)$ exactly. Subsequently, I describe another formula that can roughly approximate the probability $NA(k, r, 2N)$. In devising this formula, I use an approximate procedure considering only those events that make major contributions to $NA(k, r, 2N)$ and ignoring those that are likely to make only marginal contributions to its value. Although it is possible to obtain exact values for these probabilities simply by performing a large number of Monte Carlo simulations, analytical expressions are useful because they can help us obtain a intuitive understanding of the processes involved. Simulation experiments indicate that the probabilities calculated using my approximate formula are always comparable to the true expectations provided that the number of generations ($k$) remains small (e.g. <= 100). C++ programs that can efficiently calculate these formulas are available upon request.

## Calculating $L(k, r, 2N)$

$L(k, r, 2N)$ refers to the probability that any randomly chosen chromosome from the current population belongs to one of the 2 classes of chromosomes (see descriptions below) that don't leave behind any copies after $k$ generations. $L(k, r, 2N) = T1(k, r, 2N) + T2(k, r, 2N)$, where the 2 terms correspond to the probabilities of the following events respectively:

1) A chromosome loses all its descendants within the next $k$ generations without any of the homologous chromosomes of either the chromosome or its descendants having undergone recombination. (Class 1)

2) A chromosome's homologous chromosome recombines in the first generation but both of them lose all their descendants without any of the homologous chromosomes of their descendants having undergone any recombination in the subsequent $k - 1$ generations. (Class 2)

Let us assume that there are $n$ copies of a chromosome in the current generation and exactly $m$ copies get picked in the next generation. The probability that this happens is

$$s(n, m) = \binom{2N}{m} (n/2N)^m (1 - n/2N)^{2N - m}$$

, where $2N$ denotes the total number of chromosomes in the diploid population and $\binom{2N}{m}$ means $2N$ choose $m$.

If the population size can change between current and next generations, then a more general expression would be: $g(n, m, N_1, N_2) = \binom{N_2}{m} (n/N_1)^m (1 - n/N_1)^{N_2-m}$, where $N_1$ and $N_2$ represent the population sizes of the current and next generations.

The $n$ copies can occur as $x$ pairs of homologous chromosomes and $y$ singleton copies in independent individuals such that $2x + y = n$. The probability of such a configuration is given by $h(x, y, n) = \binom{N}{x}\binom{N-x}{n-2x} 2^{n-2x} / \binom{2N}{n}$, where $x$ can take values from $max(0, n - N)$ to $[n/2]$. Here, [ ] means integer part of.

Given $n$ copies of a chromosome in the current generation, the probability that exactly $m$ copies get picked in the next generation and none of the homologous chromosomes of those $n$ copies recombine is given by:

$sh(n, m) = g(n, m, 2N, 2N)$ x

P(Homologs of copies don't recombine | current = $n$, next = $m$)

If there are *x* pairs and *y* singleton copies in the current generation, the chance that none of the homologs of the copies recombine can be calculated as follows:

The probability that a chromosome that gets picked is non-recombinant is $e^{-r}$.

P(*y* homologs of the singleton copies don't recombine | current = *n*, next = *m*) =

$$\sum_{z=0}^{z=2N-m} g(y, z, 2N-n, 2N-m) e^{-zr}$$

P(2*x* homologs of the copies that occur in pairs don't recombine | current = *n*, next = *m* )

$$= \sum_{z=0}^{z=m} g(2x, z, n, m) e^{-zr}$$

Note that if exactly *m* copies of the chromosome get picked in the next generation, then the homologs of the singleton copies can get picked at most 2*N* – *m* times. Similarly, copies of the chromosome that are homologs of other copies of the chromosome within diploids, can get picked at most *m* times.

P(Homologs of the copies don't recombine | current = *n,* next = *m* ) =

$$\sum_{x=max(0,\ n-N)}^{x=[n/2]} [(\sum_{z=0}^{z=m} g(2x, z, n, m) e^{-zr}) \times (\sum_{z=0}^{z=2N-m} g(y, z, 2N-n, 2N-m) e^{-zr}) \times h(x, y, n)]$$

Here, we average over all the possible configurations in which the *n* copies can occur.

Thus, $sh(n, m) = g(n, m, 2N, 2N)$ x

$$\sum_{x=max(0,\ n-N)}^{x=[n/2]} [(\sum_{z=0}^{z=m} g(2x,z,n,m)e^{-zr}) \text{ x } (\sum_{z=0}^{z=2N-m} g(y,z,2N-n,2N-m)e^{-zr}) \text{ x } h(x, y, n)]$$

Let *gen(0)* represent the previous generation, *gen(1)* represent the generation being simulated and *gen(2), gen(3)* ...etc represent subsequent generations. *P1(1, a, r, 2N)*, the probability that all the descendants of a chromosome from *gen(1)* are lost at *gen(a+2)* but not before that is given by:

$P1(1, a, r, 2N) =$

$$\sum_{m1=1}^{m1=2N-1} \sum_{m2=1}^{m2=2N-1} ... \sum_{ma=1}^{ma=2N-1} sh(m0,m1)sh(m1,m2)...sh(ma-1,ma)sh(ma,0)$$

where $m0 = 1$ initially.

Note that *P1(1, a, r, 2N)* is simply the probability that a *time-homogeneous markov chain* starting from state 1 (copies) transitions to state 0 (copies) for the first time after exactly $a + 1$ steps, where the transition probabilities at each step are given by *sh(n, m)*. The total probability *T1(k, r, 2N)* that all the descendants of a chromosome from *gen(1)* are lost by *gen(k +1)* without any of their homologs having undergone recombination is

therefore equal to: $\sum_{a=0}^{a=k-1} P1(1,a,r,2N)$

Similarly, if the homolog of a chromosome recombines in the first generation, the chance of transition to a total of $m$ copies for the chromosome and its homolog is given by:

$$ss(2, m) = \sum_{z=0}^{m} s(1,z)g(1, m-z, 2N-1, 2N-z)(1-e^{-(m-z)r})$$

Here, $z$ is the number of times the chromosome gets picked and $m - z$ is the number of times its homolog gets picked. $1 - e^{-(m-z)r}$ is the chance that at least one of the copies of the homolog that gets picked, undergoes recombination.

$P2(2, a, r, 2N)$, the probability that the homolog of a chromosome recombines in the first generation and all the descendants of both the chromosome and homolog disappear at $gen(a+2)$ but not before that is given by:

$$P2(2,a,r,2N) = \sum_{m1=1}^{m1=2N-1} \sum_{m2=1}^{m2=2N-1} \cdots \sum_{ma=1}^{ma=2N-1} ss(2,m_1)sh(m_1,m_2)\ldots sh(m_{a-1},m_a)sh(m_a,0)$$

Note that $P2(2,a,r,2N)$ is simply the probability that a *time-inhomogeneous markov chain* starting from state 2 (copies) transitions to state 0 (copies) for the first time after exactly $a + 1$ steps, where the transition probabilities for the first step is $ss(2, m)$ and for any subsequent step is $sh(n, m)$.

The total probability that the homolog of a chromosome recombines in the first generation and all the descendants of both a chromosome and its homolog disappear by *gen(k + 1)* without the homologs of the descendants undergoing any recombination in the last *k – 1* generations is then given by:

$$T2(k, r, 2N) = \sum_{a=0}^{a=k-1} P2(2,a,r,2N)$$

Thus, *L(k, r, 2N)* =

$$T1(k, r, 2N) + T2(k, r, 2N) = \sum_{a=0}^{a=k-1} P1(1,a,r,2N) + \sum_{a=0}^{a=k-1} P2(2,a,r,2N)$$

Table 1 shows comparisons between simulations and the formula given above.

**Table 1**

**Probability *L(k, r, 2N)* calculated from formula and from 100 million simulations**

| k | r | 2N = 100 | | 2N = 1000 | |
|---|---|---|---|---|---|
| | | *Simulated* | *Formula* | *Simulated* | *Formula* |
| 2 | 0.00 | 0.529055 | 0.529053 | 0.531224 | 0.531224 |
| | 0.25 | 0.408460 | 0.408454 | 0.411124 | 0.411125 |
| | 1.00 | 0.257541 | 0.257543 | 0.260652 | 0.260652 |
| | 10.0 | 0.173947 | 0.173944 | 0.177065 | 0.177066 |
| 4 | 0.00 | 0.685091 | 0.685090 | 0.687639 | 0.687639 |
| | 0.25 | 0.483721 | 0.483715 | 0.486706 | 0.486708 |
| | 1.00 | 0.278997 | 0.279001 | 0.282407 | 0.282407 |
| | 10.0 | 0.181926 | 0.181930 | 0185274 | 0.185275 |
| 8 | 0.00 | 0.807862 | 0.807856 | 0.810647 | 0.810644 |
| | 0.25 | 0.503710 | 0.503712 | 0.506713 | 0.506714 |
| | 1.00 | 0.280459 | 0.280458 | 0.283887 | 0.283887 |
| | 10.0 | 0.182135 | 0.182135 | 0.185487 | 0.185488 |

**Bounds for *NA(k, r, 2N)***

*L(k, r, 2N)* is a lower bound for *NA(k, r, 2N)* because a chromosome that does not belong to either of the 2 classes we described before, can still disappear without leaving a trace.

To establish that *L(k, 0, 2N)* is an upper bound, let $A_i$ denote the event that the $i^{th}$ base pair of a chromosome does not leave copies after *k* generations.

Then, *NA(k, r, 2N)* = Probability($A_1 \cap A_2 \cap A_3 \ldots A_L$), where *L* denotes the total length of a chromosome in base pairs and $\cap$ means intersection.

When there is no recombination, it is evident that:

Probability($A_1$) = Probability($A_2$) = Probability($A_3$)… = Probability($A_L$) = *L(k, 0, 2N)*

and the events $A_i$ are all identical. Note that when we focus on a single basepair in a chromosome, it will be inherited from a single parental chromosome alone. Thus, the Wright-Fisher model with recombination reduces to the Wright-Fisher model without recombination (Hein et al 2005). Hence, the probability of event $A_i$ is independent of the recombination rate *r*.

Probability($A_1 \cap A_2 \cap A_3 \ldots\ldots A_L$) <= Probability($A_i$) = *L(k, 0, 2N)* since the events $A_i$ are not identical if there is recombination. So, *L(k, r, 2N)* <= *NA(k, r, 2N)* <= *L(k, 0, 2N)*.

**An approximate formula for *NA(k, r, 2N)***

*NA(k, r, 2N)* refers to the probability that a chromosome in the current population does not leave any copies in the future population that exists after *k* generations. We consider a single chromosome in generation 0, which we will refer to as the initial chromosome. In guessing an approximate expression for *NA(k, r, 2N),* I will assume that all the copies of a chromosome that get picked will always occur within different individuals in the population and hence recombinations will always happen between a copy and a noncopy. The intuition behind this assumption is that, since we initially start with a single copy, the number of copies of the chromosome is unlikely to rise to high numbers in the short-term (i.e. when *k* is small) if the population size is large enough. Therefore, the chance that some of the copies will occur within the same individuals is small.

First, we classify the copies of a chromosome in the current generation (*l*) into 3 types. Let *m1* denote the number of copies of a chromosome that have not undergone any recombination events in their history (non-recombinant copies), *m2* denote the number of copies that have undergone exactly 1 recombination event and *m3* denote copies that have undergone more than 1 recombination events in their history. History here refers to the events that have happened as we follow the ancestors (only those that are also copies of the initial chromosome) of a copy into the past until we reach the initial chromosome in generation 0.

The chance that we will transition to *m4* non-recombinant copies, *m5* copies with exactly 1 recombination event in their history and *m6* copies with more than 1 recombination in generation *l + 1* is calculated using a multinomial distribution:

*T(m1, m2, m3, m4, m5, m6, l, 2N)*

$= 2N! p_1^{m4} p_2^{m5} p_3^{m6} (1 - p_1 - p_2 - p_3)^{2N - m4 - m5 - m6} / m4! m5! m6! (2N - m4 - m5 - m6)!$

where *2N* denotes the total number of chromosomes in the diploid population and:

$p1 = m1 e^{-r}/2N$

$p2 = (2m1(1 - e^{-r})/2N) + (m2 e^{-r}/2N)$

$p3 = (1.5 m2(1 - e^{-r})/2N) + (m3 e^{-r}/2N) + (m3 d(l)(1 - e^{-r})/2N)$

Here $d(l) = 1 + \sum_{x=2}^{x=l} f1(x) p(x, l)$ where the terms can be calculated recursively as:

$f1(x) = 0.5 f1(x - 1)/f2(x - 1)$ and $f2(x) = 0.5 + 0.5 f1(x)$ with $f1(0) = 1.0$ initially.

$p(x, l) = [p(x, l - 1) e^{-r} + 2p(x - 1, l - 1)(1 - e^{-r}) f2(x - 1)]/s$ and

$s = \sum_{x=2}^{x=l} [p(x, l - 1) e^{-r} + 2p(x - 1, l - 1)(1 - e^{-r}) f2(x - 1)]$ with $p(2, 2) = 1.0$ initially.

We model the change in the number of copies of the 3 types (i.e. from ($m1, m2, m3$) to ($m4, m5, m6$)) at each generation as a *Markov Chain* with the transition probabilities described previously. Note that the transition probabilities depend on the number of generations and thus this represents a *time-inhomogeneous chain*. I propose the following approximation:

$$NA(k, r, 2N) \approx \sum_{m1=0}^{m1=2N} \sum_{m2=0}^{m2=2N-m1} \sum_{m3=0}^{m3=2N-m1-m2} f(m1, m2, m3, k-1, 2N) T(m1, m2, m3, 0, 0, 0, k-1, 2N)$$

Here $f(m1, m2, m3, l, 2N)$ denotes the probability of observing $m1$, $m2$ and $m3$ copies of the 3 types in the $l^{th}$ generation. These probabilities can be calculated recursively using:

$$f(m1, m2, m3, l, 2N) = \sum_{m4=0}^{m4=2N} \sum_{m5=0}^{m5=2N-m4} \sum_{m6=0}^{m6=2N-m4-m5} f(m4, m5, m6, l-1, 2N) T(m4, m5, m6, m1, m2, m3, l-1, 2N)$$

with

$f(1, 0, 0, 0, 2N) = 1.0$ at the start and the rest of the $f$ probability values 0 in the $0^{th}$ generation (i.e. initially we have a single copy of a non-recombinant chromosome in the population).

Note that $NA(k, r, 2N)$ is the probability that a *time-inhomogenous markov chain* with transition probability of $T(m_1, m_2, m_3, m_4, m_5, m_6, l, 2N)$ transitions from an initial state of $(1, 0, 0)$ to a state of $(0, 0, 0)$ within the first $k$ steps. In the following 2 sections, I will describe the rationale for choosing the various terms described in this approximate formula. The key idea is to consider only the average scenarios at each step and this makes calculations computationally tractable and yields a formula that is relatively simple.

**Descriptions of the terms in the formula**

$p_1$ denotes the probability that a particular draw from the current generation ($l$) creates a non-recombinant copy in the new generation. This is equal to the chance that a non-recombinant copy is picked ($= m_1/2N$) multiplied by the chance that it does not recombine in the current generation ($= e^{-r}$).

$p_2$ denotes the probability that a particular draw from the current generation ($l$) creates a copy in the new generation that has had exactly 1 recombination in its history. This can happen if a non-recombinant copy or its homolog is picked (probability $= 2m_1/2N$) and it recombines in the current generation (probability $= 1 - e^{-r}$). The probability of this event $= 2m_1(1 - e^{-r})/2N$. Such a copy can also be created if a copy that has had exactly 1 recombination event so far is picked, and it does not recombine in the current generation.

The probability of this event is = $(m2/2N) \times e^{-r}$.

$p_3$ denotes the probability that a particular draw from the current generation (*l*) creates a copy in the new generation that has had more than 1 recombination event in its history.

i) This can happen if a copy that has had exactly 1 recombination so far or its homolog gets picked (probability = $2m2/2N$) and recombines in the current generation (probability = $1 - e^{-r}$) such that the recombinant carries material from the initial chromosome. The probability of this event is $< 2m2(1 - e^{-r})/2N$. This is because, on average, a copy with exactly 1 recombination in its history, only carries 50% material from the initial chromosome. Without loss of generality, we can assume that the top 50% of a copy is the material from the initial chromosome. If the recombinant is created after picking the copy, then it will always carry some material from the initial chromosome (see description at start). If the homolog of the copy gets picked (i.e. a non-copy), then the chance that the recombinant carries material from initial chromosome = 0.5 i.e. only when the breakpoint falls within the material from the initial chromosome. Therefore, the chance that the new copy will carry material from the initial chromosome is equal to $(1 + 0.5) \times m2(1 - e^{-r})/2N$ for the average case.

ii) A copy of third type can also be created if a chromosome copy that has had more than 1 recombination event so far is picked, and it does not recombine in the current generation. The probability of this event is = $m3/2N \times e^{-r}$.

iii) Lastly, copies that have had more than 1 recombination event so far or their homologs can get picked and recombine in the current generation to create copies of the third type. The chance of these events is given by $m3d(l)(1 - e^{-r})/2N$. Here, $d(l)$ denotes a term that accounts for the average decrease in the material from the initial chromosome in the copies that have undergone more than 1 recombination event as the number of generations ($l$) increases. As $l$ increases, we expect the average material from the initial chromosome to decrease since the expected number of recombination events increases.

**Calculating *d(l)***

We calculate *d(l)* using an iterative procedure considering the average cases at each step.

Assuming that exactly $x$ recombination events have happened in the history of a copy, let *f1(x)* denote the fraction of the chromosome that comes from the initial chromosome (for the average case) and *f2(x)* denote the chance that a recombination between such a copy and its homolog produces another copy.

For example, initially $x = 0$ and so *f1(x)* = 1.0 and *f2(x)* = 1.0.

For $x = 1$, *f1(x)* = 0.5 and *f2(x)* for the average case = 0.5 + 0.5*f1(x)* = 0.75.

For $x = 2$, we only consider the average case from $x = 1$ i.e. when *f1(x - 1)* = 0.5. Now, if the fraction of material in a copy is 0.5, we can calculate the expectation of the fraction after 1 more recombination.

Note that a recombinant chromosome can be created either after choosing a copy or its homolog (a noncopy). Without loss of generality, we can also assume that the topmost fraction *f1(x - 1)* of the copy carries the material from the initial chromosome. 4 different cases arise when creating a recombinant:

i) When the copy gets chosen and recombines, the chance that the fraction of material in the recombinant is *f1(x – 1)* = 1 - *f1(x – 1)* (When the break point is beyond the end of material from initial chromosome).  …case 1

ii) When the breakpoint falls within the material from initial chromosome, the average fraction in recombinant is $0.5f1(x-1)$ since recombination is uniform. Chance that the breakpoint falls within the material from initial chromosome $= f1(x-1)$.     …case 2

iii) When the homolog of a copy gets chosen, the chance that breakpoint falls within material from initial chromosome is also $= f1(x-1)$ and the fraction of material in the recombinant is on average $0.5f1(x-1)$.     …case 3

iv) If the homolog gets picked and the breakpoint falls outside the material from the initial chromosome, the resultant recombinant does not carry any material from the initial chromosome.     …case 4

$f1(x)$ = Average fraction of material in a copy with $x$ recombinations in its history. The chance that a recombinant is created after picking copy $= 0.5$ and chance that recombinant gets created by picking its homolog $= 0.5$.

The chance that the recombinant carries material from initial chromosome $f2(x-1)$
$= 0.5(1 - f1(x-1) + f1(x-1) + f1(x-1))$ (the 3 terms correspond to cases 1 to 3).

The average value of $f1(x)$ given that the recombinant is a copy =

$f1(x-1)[(0.5(1 - f1(x-1)))/f2(x-1)] \quad + \quad 0.5\,f1(x-1)[0.5\,f1(x-1)/f2(x-1)]$

$+\, 0.5\,f1(x-1)[0.5\,f1(x-1)/f2(x-1)]$ (The 3 terms correspond to cases 1 to 3).

Thus, $f1(x) = 0.5 f1(x-1)/f2(x-1)$ and $f2(x) = 0.5 + 0.5 f1(x)$ with initial condition of $f1(0) = 1.0$.

Next, we account for the relative frequencies of the copies with different values of $x$ in the $l^{th}$ generation (denoted by $p(x, l)$). This can be done as follows for the copies of the third type.

In the $2^{nd}$ generation, all the copies of the third type will have exactly 2 recombination events in their history. Thus, $p(2, 2) = 1.0$ and 0 for any other value of $x$.

Now, in the $3^{rd}$ generation, it is easy to see that:

chance of $x = 2$ is proportional to $p(2, 2) \, e^{-r}$

i.e. copies with 2 recombinations don't recombine.

chance of $x = 3$ is proportional to $p(3, 2) \, e^{-r} + 2p(2, 2)(1 - e^{-r}) f2(2)$

i.e. copies with 3 recombinations don't recombine or those with 2 recombinations (or their homologs) recombine in current generation and the recombinant carries material from the initial chromosome.

In general, it is easy to see that in the $l^{th}$ generation:

$p(x, l)$ is proportional to $p(x, l - 1) \, e^{-r} + 2p(x - 1, l - 1)(1 - e^{-r}) f2(x - 1)$

Let $s = \sum_{x=2}^{x=l} p(x, l - 1) \, e^{-r} + 2p(x - 1, l - 1)(1 - e^{-r}) f2(x - 1)$

Then, $p(x, l) = [p(x, l - 1) \, e^{-r} + 2p(x - 1, l - 1)(1 - e^{-r}) f2(x - 1)]/s$ with initial condition of $p(2, 2) = 1.0$.

Once we have $p(x, l)$, we can calculate the average fraction of material from the initial chromosome in copies of the third type. This is equal to $\sum_{x=2}^{x=l} f1(x)p(x, l)$

Without loss of generality, we can assume that this material occurs at the topmost in a copy. If the recombinant is created after choosing a copy, then it will always carry some material from the initial chromosome (see model description in Introduction). If the homolog of the copy gets chosen (i.e. a non-copy), then the chance that the recombinant carries material from initial chromosome = $\sum_{x=2}^{x=l} f1(x)p(x, l)$ on average (i.e. only when the breakpoint falls within the material from the initial chromosome).

Let $d(l) = (1 + \sum_{x=2}^{x=l} f1(x)p(x, l))$. Thus, the average chance that recombinants involve chromosomes of the third type and carry material from the initial chromosome is given by

$m3(1 + \sum_{x=2}^{x=l} f1(x)p(x, l))(1 - e^{-r})/2N$. $(1 - e^{-r})$ is the probability of recombination and

$m3$ denotes the total number of copies of the third type in the current generation.

$m3(1 - e^{-r})/2N$ denotes the chance that the recombinant is created after picking the copies and $m3 \sum_{x=2}^{x=l} f1(x)p(x, l)(1 - e^{-r})/2N$ denotes the chance that the recombinant is created after picking their homologs (see description of model in Introduction).

Note that in the formula described previously, I am not attempting to calculate the correct probability of loss (which is tedious and computationally expensive) but instead modeling a scenario where the average cases always occur in each generation. In practice, I truncate the summations in this formula after a fixed number of terms to obtain rough approximations for the value of *NA(k, r, 2N)*. Thus, I use approximate formulas of the type:

$$NA(k, r, 2N) \approx \sum_{m1=0}^{m1=t1} \sum_{m2=0}^{m2=t2} \sum_{m3=0}^{m3=t3} f(m1, m2, m3, k-1, 2N) T(m1, m2, m3, 0, 0, 0, k-1, 2N)$$

where *f(m1, m2, m3, l, 2N)* =

$$\sum_{m4=0}^{m4=t1} \sum_{m5=0}^{m5=t2} \sum_{m6=0}^{m6=t3} f(m4, m5, m6, l-1, 2N) T(m4, m5, m6, m1, m2, m3, l-1, 2N)$$

and *f(1, 0, 0, 0, 2N)* = 1.0 at the start and the rest of the *f* probability values are 0 in the $0^{th}$ generation. The number of terms *t1*, *t2* and *t3* can be chosen empirically. Note that the proposed approximation is both relatively simple as well as flexible. More generally, the number of terms in the formula (*t1(l), t2(l), t3(l)*) as well as the choice of an expression for *d(l)* can also be varied as a function of the number of generations (*l*). Figure 1 shows comparisons between my approximation and simulations. The primary mathematical insight we derive from the closeness of the simulated and analytical curves is that the probability of loss in the presence of recombination is determined mainly by the total amount of material from the initial chromosome occurring in a generation (on average) whereas the details of how the material is distributed across chromosomes as well as how individual recombination events are taking place in each generation are less important.

**Figure 1**

*NA(k, r, 2N)* as calculated from my approximate formula with t1 = 23, t2 = 29 and t3 = 37 and from 100, 000 simulations.

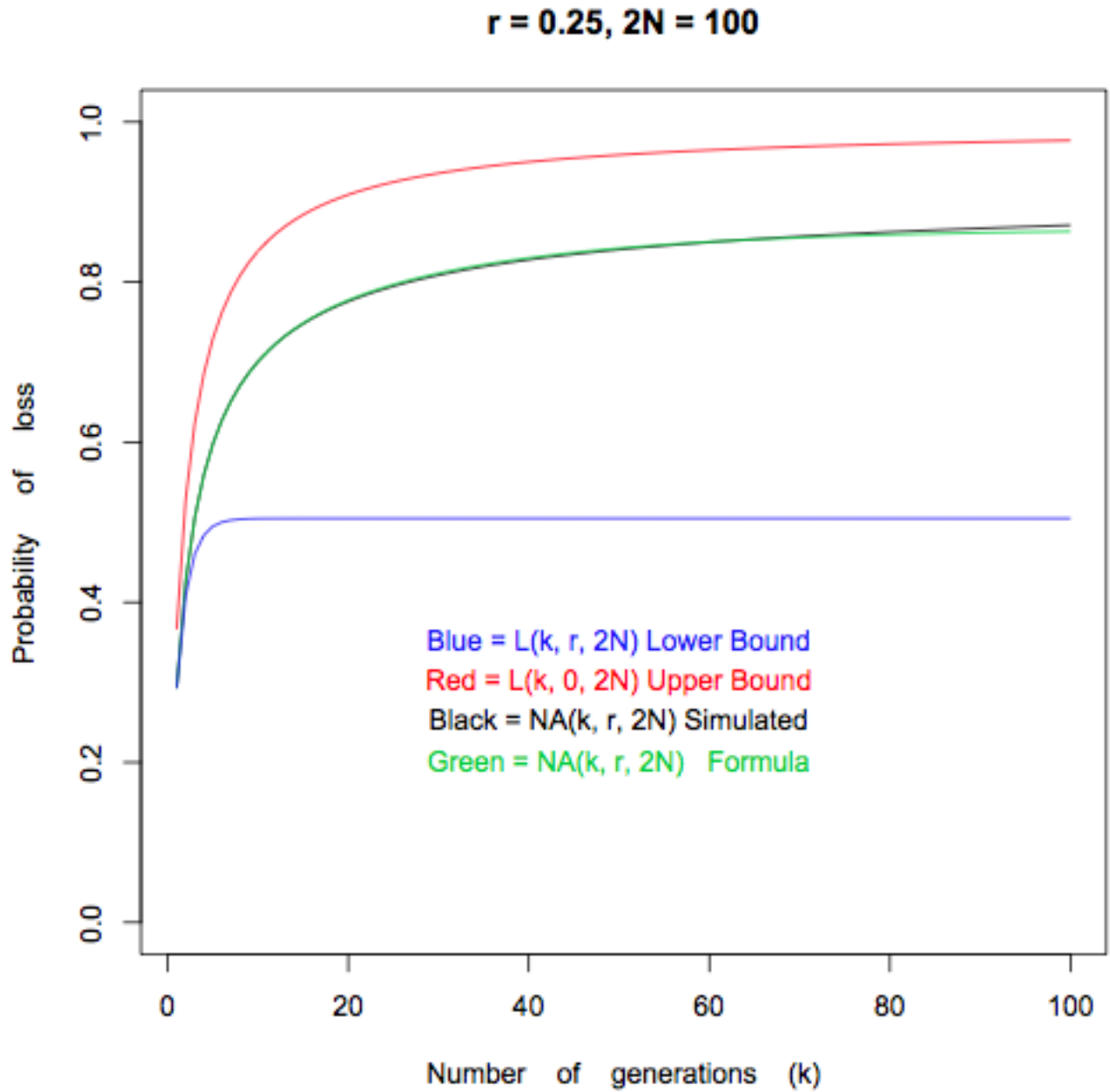

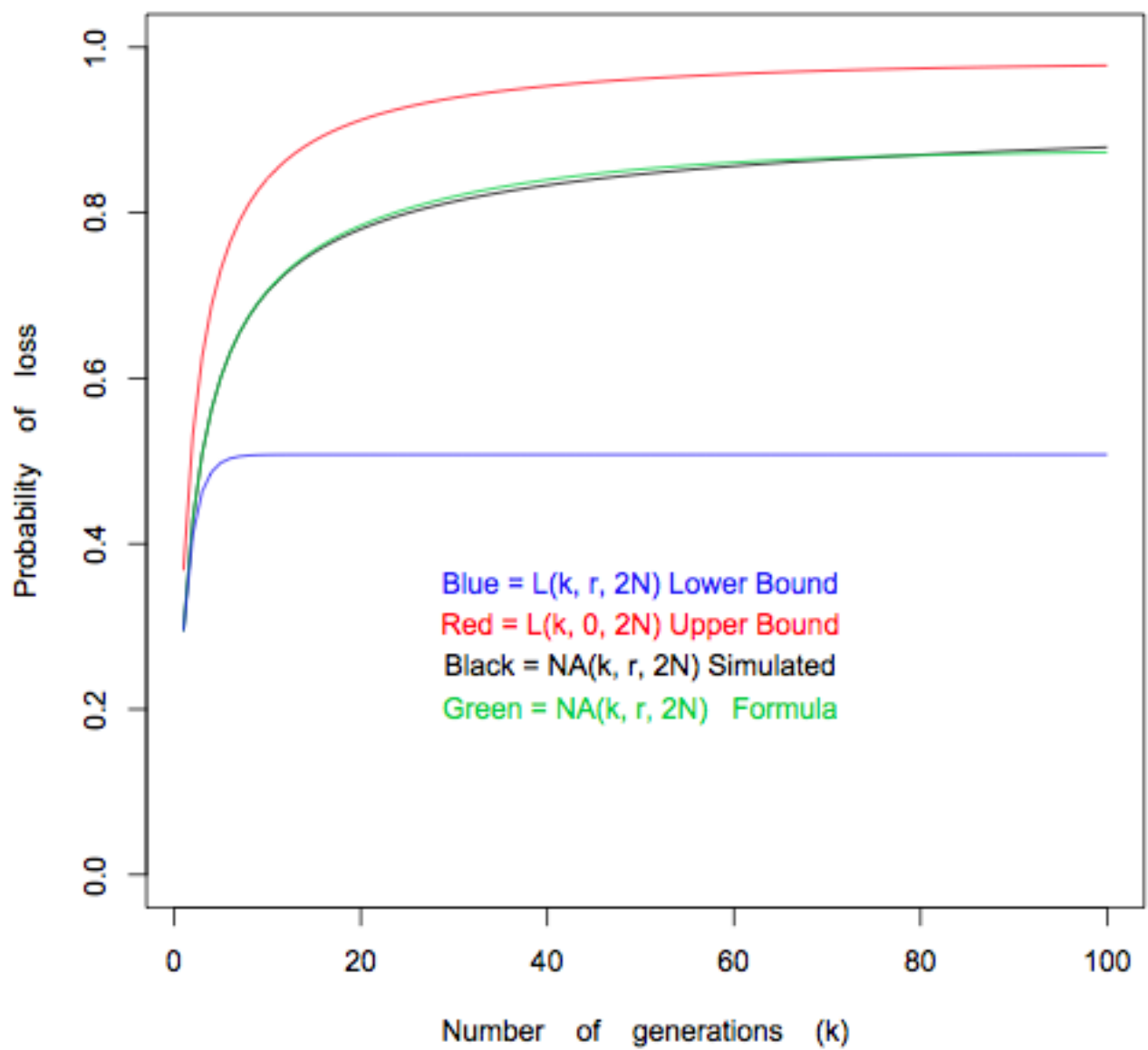

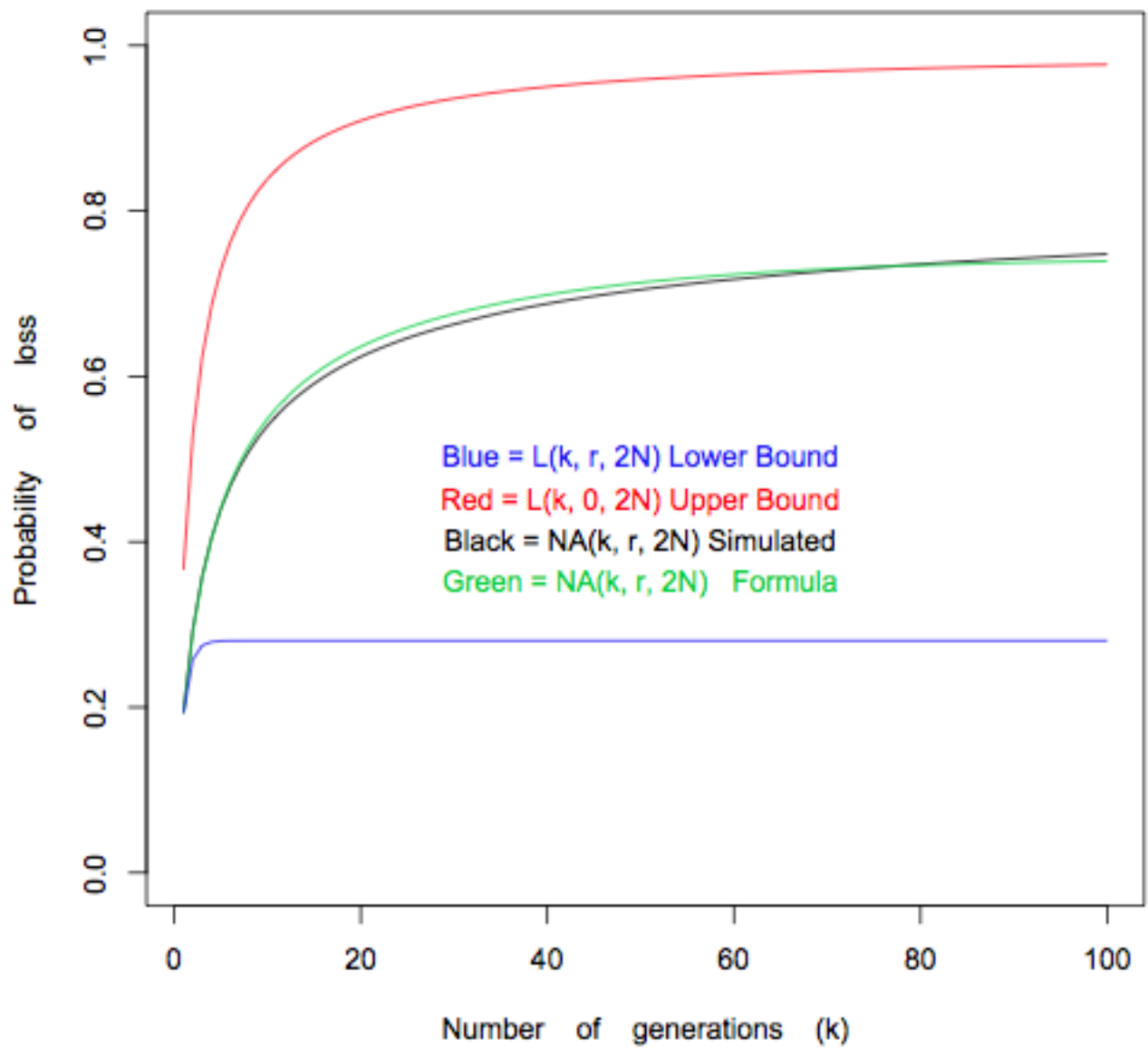

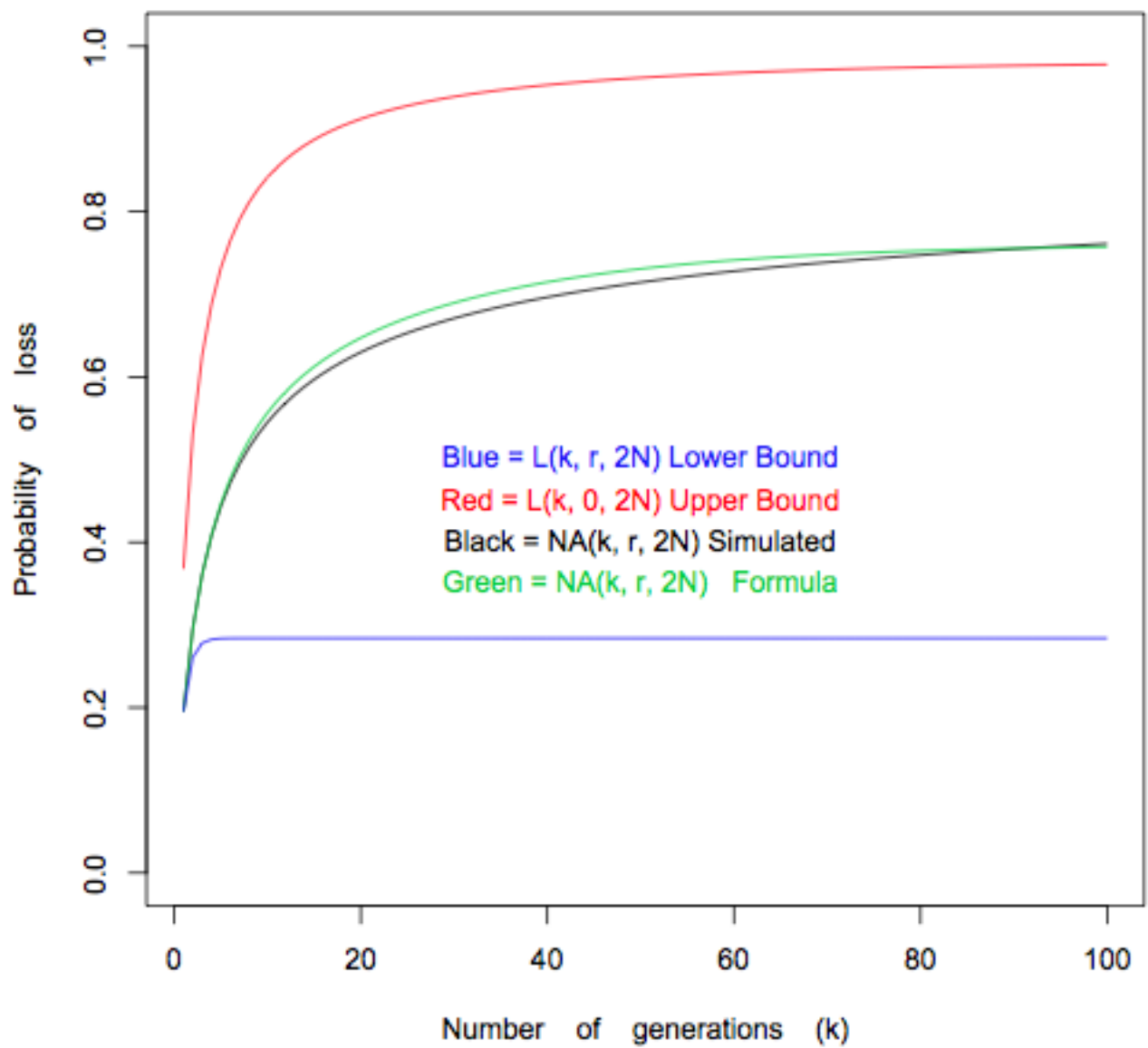

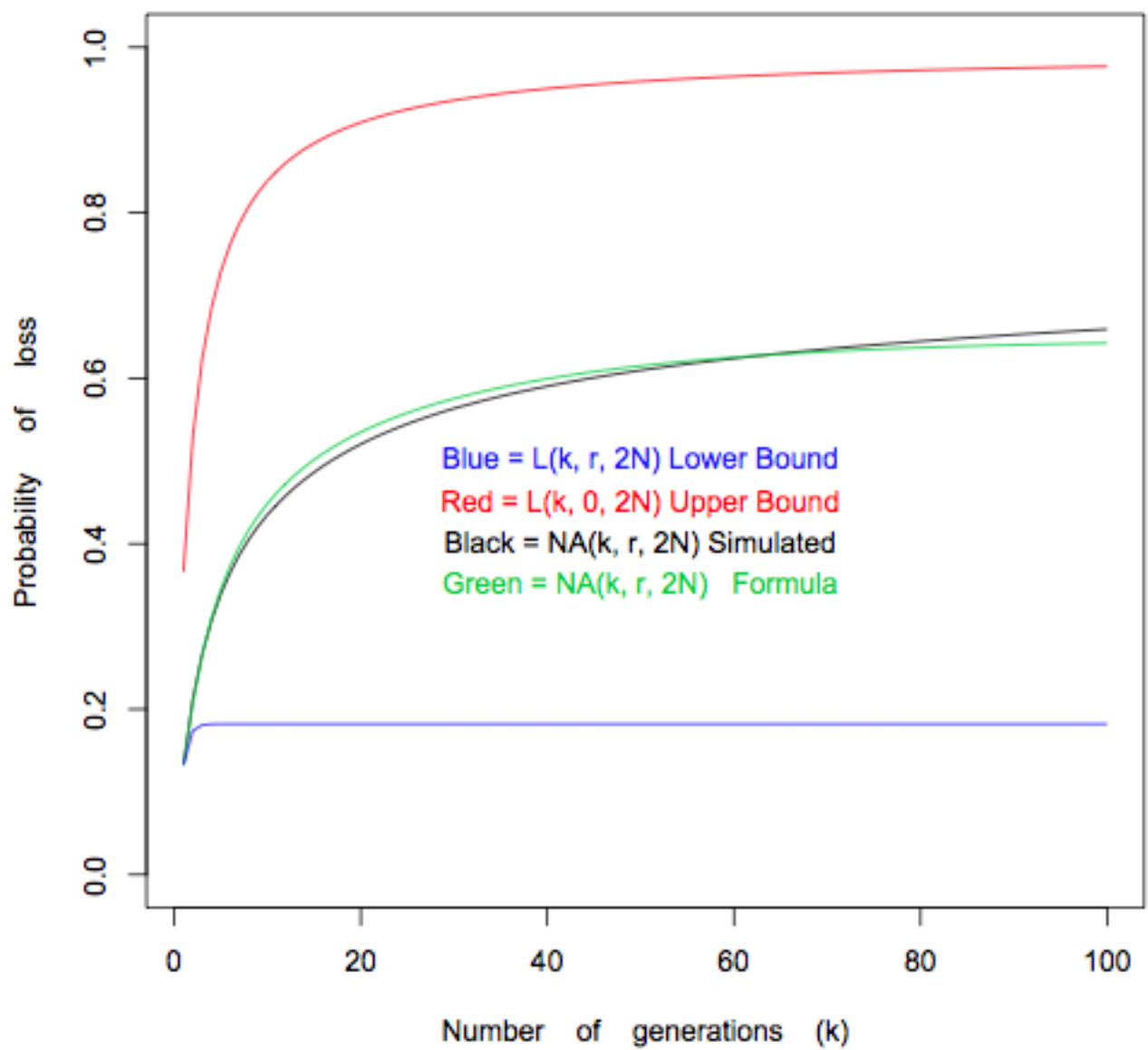

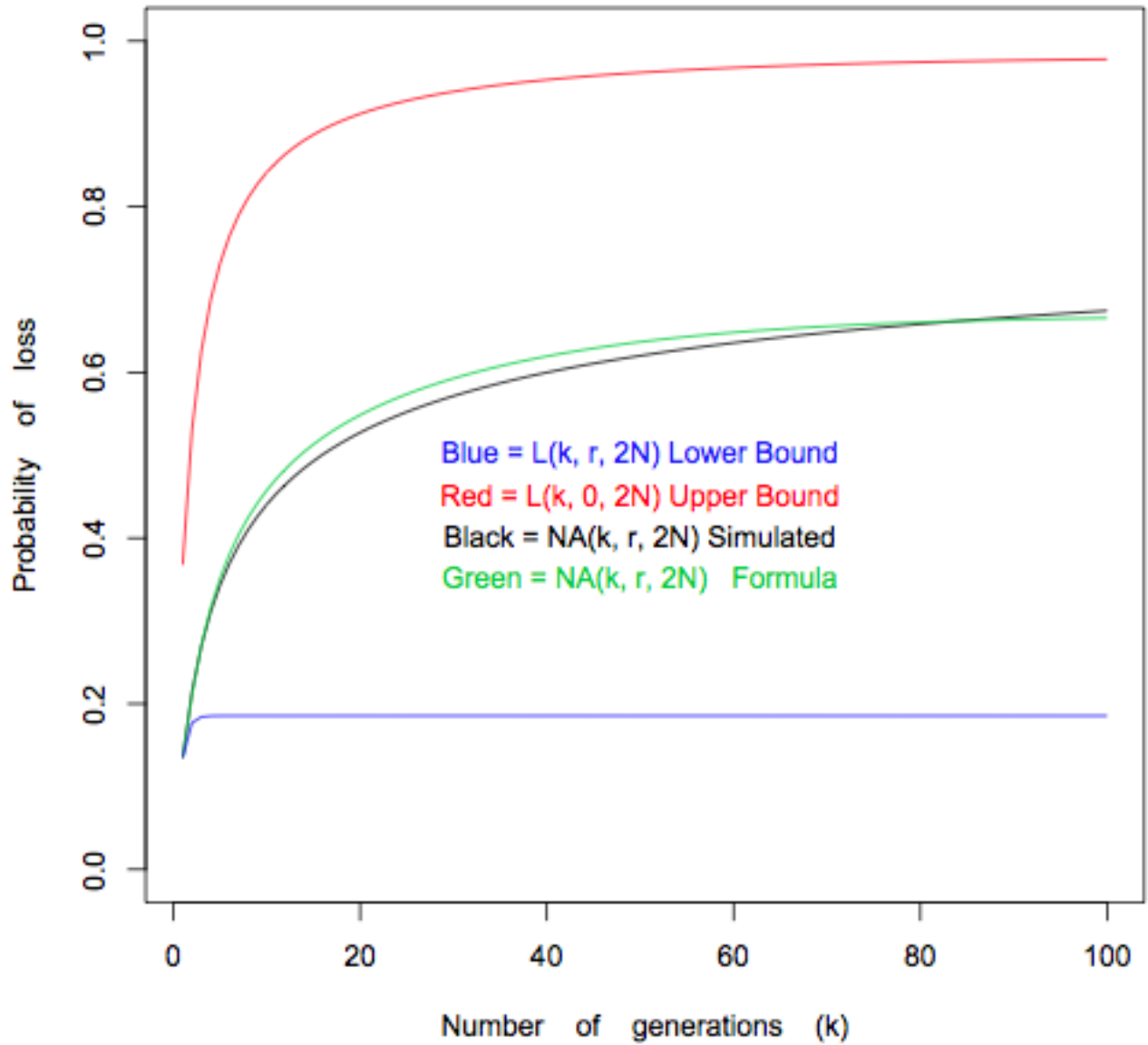

**Discussion and Conclusion**

I have presented a computationally tractable approximate formula for calculating the short-term probability of loss of a chromosome under the neutral Wright-Fisher model in the presence of recombination. In the forward simulation algorithm that was described in Padhukasahasram et al. 2008, we use a look-ahead procedure in order to avoid performing operations on non-ancestral chromosomes during the course of simulating a Wright-Fisher population. This procedure improves the run-time efficiency and also reduces the memory used during the run. The formulas described in this paper can help us determine the fraction of chromosomes in the current population that can be skipped, when we look ahead for a fixed number of generations into the future for a given recombination rate. Note that the expected value of this fraction as calculated from simulations is equal to the probability of loss. Since our simulation program (FORWSIM) makes a trade off between the computational effort required to look ahead and identify non-ancestral chromosomes from the current generation and the effort saved by skipping operations on such chromosomes, these probability formulas can guide the choice of the look-ahead parameter (i.e. number of generations) when different variations of our forward algorithm are implemented (for e.g. we could use different look-ahead strategies as well as parameters at different stages of the simulation, see Padhukasahasram et al. 2008).

**Acknowledgements**

This research project was supported by NHGRI grant HG01988 to Dr. Bruce Rannala.

# References

Hoggart CJ, Chadeau-Hyam M, Clark TG, Lampariello R, Whittaker JC, De Iorio M, Balding DJ (2007) Sequence-level population simulations over large genomic regions. *Genetics* 177: 1725-1731

Durret, R (2002) *Probability models for DNA sequence evolution.* Springer, New York

Fisher, RA (1930) *The genetical theory of natural selection.* Clarendon Press, Oxford

Griffiths RC (1981) Neutral two-locus multiple allele models with recombination. *Theor Popul Biol* 19: 169-186

Hein, J, Schierup, MH, and Wiuf, C (2005) *Gene genealogies, variation and evolution: a primer in coalescent theory.* Oxford University Press

Hernandez RD (2008) A flexible forward simulator for populations subject to selection and demography. *Bioinformatics* 24: 2786-2787

Hudson, RR (1983) Properties of a neutral allele model with intragenic recombination. *Theor Popul Biol* 23: 183-201

Kimmel, M and Peng, B (2005) simuPOP: a forward-time population genetics simulation environment. *Bioinformatics* 21: 3686-3687

Kimura, M and Ohta, T (2001) *Theoretical aspects of population genetics.* Princeton University Press

Kingman, JFC (1982) On the genealogy of large populations. *Stoch Proc Appl* 13: 235-248

Padhukasahasram, B, Marjoram, P, Wall, JD, Bustamante, CD and Nordborg, M (2008) Exploring population genetic models with recombination using efficient forward-time simulations. *Genetics* 178**:** 2417–2427

Wakeley, J (2008) *Coalescent theory.* Ben Roberts, Greenwood Village

Wright, S (1931) Evolution in Mendelian populations. *Genetics* 16: 97-159